\documentclass
[superscriptaddress,secnumarabic,amssymb,amsmath,nobibnotes,aps,prd,showkeys,showpacs,twocolumn,nofootinbib]{revtex4}%
\usepackage{graphicx}
\usepackage{epsf}
\usepackage{bm}
\usepackage{amsmath}
\usepackage{amsfonts}
\usepackage{amssymb}
\usepackage{epstopdf}%
\providecommand{\U}[1]{\protect\rule{.1in}{.1in}}
\newcommand{\be}{\begin{equation}}
\newcommand{\ee}{\end{equation}}

\newcommand{\mincir}{\raise
-3.truept\hbox{\rlap{\hbox{$\sim$}}\raise4.truept\hbox{$<$}\ }}
\newcommand{\magcir}{\raise
-3.truept\hbox{\rlap{\hbox{$\sim$}}\raise4.truept\hbox{$>$}\ }}

\begin{document}
\title{Invariant solutions and Noether symmetries  
in Hybrid  Gravity}

\author{Andrzej Borowiec}
\email{andrzej.borowiec@ift.uni.wroc.pl}
\affiliation{Institute for Theoretical Physics, pl. M. Borna 9, 50-204,
Wroclaw, Poland.}

\author{Salvatore Capozziello}
\email{capozzie@na.infn.it}
\affiliation{Dipartimento di Fisica, Universit\`{a} di Napoli ``Federico II'', Napoli, Italy.}
\affiliation{Istituto Nazionale di Fisica Nucleare (INFN) Sez. di Napoli, Compl. Univ. di Monte S. Angelo, Edificio G, Via Cinthia, I-80126, Napoli, Italy.}
\affiliation{Gran Sasso Science Institute (INFN), Via F. Crispi 7, I-67100, L' Aquila, Italy.}

\author{Mariafelicia De Laurentis}
\email{mfdelaurentis@tspu.edu.ru}
\affiliation{Tomsk State Pedagogical University, 634061 Tomsk and National Research Tomsk State University, 634050 Tomsk, Russia.}
\affiliation{Dipartimento di Fisica, Universit\`{a} di Napoli ``Federico II'', Napoli, Italy.}
\affiliation{Istituto Nazionale di Fisica Nucleare (INFN) Sez. di Napoli, Compl. Univ. di Monte S. Angelo, Edificio G, Via Cinthia, I-80126, Napoli, Italy.}

\author{Francisco S. N. Lobo}
\email{flobo@cii.fc.ul.pt}\affiliation{Centro de Astronomia
e Astrof\'{\i}sica da Universidade de Lisboa, Campo Grande, Edif\'{i}cio C8
1749-016 Lisboa, Portugal.}

\author{Andronikos Paliathanasis}
\email{paliathanasis@na.infn.it}
\affiliation{Dipartimento di Fisica, Universit\`{a} di Napoli ``Federico II'', Napoli, Italy.}
\affiliation{Istituto Nazionale di Fisica Nucleare (INFN) Sez. di Napoli, Compl. Univ. di Monte S. Angelo, Edificio G, Via Cinthia, I-80126, Napoli, Italy.}

\author{Mariacristina Paolella}
\email{paolella@na.infn.it}
\affiliation{Dipartimento di Fisica, Universit\`{a} di Napoli ``Federico II'', Napoli, Italy.}
\affiliation{Istituto Nazionale di Fisica Nucleare (INFN) Sez. di Napoli, Compl. Univ. di Monte S. Angelo, Edificio G, Via Cinthia, I-80126, Napoli, Italy.}

\author{Aneta  Wojnar}
\email{aneta.wojnar@ift.uni.wroc.pl}
\affiliation{Institute for Theoretical Physics, pl. M. Borna 9, 50-204,
Wroclaw, Poland.}


\date{\today}

\begin{abstract}

Symmetries play a crucial  role in physics and, in particular, the Noether symmetries are a useful tool both to select models motivated at a fundamental level, and to find exact solutions for  specific Lagrangians.
In this work, we consider the application of point symmetries in the recently proposed  metric-Palatini Hybrid
Gravity in order to select the $f({\cal R})$ functional form and to find analytical solutions for the field equations and for the related  Wheeler-DeWitt (WDW) equation. We show that, in order to find  out integrable  $f({\cal R})$ models,  conformal transformations in the Lagrangians  are extremely useful. In this context, we explore two conformal transformations of the forms $d\tau=N(a) dt$ and $d\tau=N(\phi) dt$. For the former conformal transformation, we found two cases of  $f({\cal R})$ functions where the field equations admit Noether symmetries. In the second case, the Lagrangian  reduces to a Brans-Dicke-like theory with a general coupling function. For each case, it is possible to transform the field equations by using  normal coordinates to simplify the dynamical system and to obtain exact solutions. Furthermore, we  perform quantization and derive  the  WDW equation for the minisuperspace model. The Lie point symmetries for the WDW equation are determined  and used to find invariant solutions. 

\end{abstract}

\maketitle

\section{Introduction}

Standard General Relativity (GR) is not enough to describe phenomena that  
recently appeared in fundamental physics, in astrophysics and cosmology such as the late-time cosmic acceleration \cite{1,2,3}, the dark matter phenomena \cite{annalen}, and several issues related to quantum field theories in curved spacetime \cite{report}. 
A  pioneering work to address  the late time acceleration  is \cite{wetterich},  where it is proposed that a fraction of the total energy density
appears  in the form of homogeneous "dark energy" and it is discussed the dynamics
which leads to the time evolution of such a dark energy.

Furthermore, modifications and extensions of the theory could be
necessary in order to explain those puzzles. 

One of the attempts is a
modification of the Hilbert-Einstein Lagrangian as a general function of the Ricci
scalar $R$, the so-called $f(R)$ theories of gravity \cite{report,sergei}. In this perspective, the dynamics of gravitation is
given by field equations which can be considered in two different approaches:
the metric and Palatini formalism. The former approach relies on the usual variation of
the action with respect to the metric tensor whereas the Palatini formalism deals with
 metric and  (torsionless) connection as two independent quantities and the
variation is taken with respect to both. In the case of GR, the two approaches
are equivalent (they provide the same field equations). However, in  $f(R)$ gravity, the metric formalism leads to a system of fourth-order partial
differential equations while the Palatini one gives second-order equations. Note also that both formalisms are dynamically equivalent to different classes of Brans-Dicke-like theories, which implies that they cannot be equivalent to each other \cite{olmo}. Thus, the $f(R)$ theories of gravity in the metric and in the Palatini formalism yield different physical predictions.

Both formalisms have been recently combined in  a new approach denoted hybrid metric-Palatini gravity
or $f(X)$ gravity \cite{Harko:2011nh,Capoz}. Here, the action is taken as the
standard Hilbert-Einstein (linear in Ricci scalar $R$) plus a non-linear term
in the Palatini curvature scalar $\cal{R}$. Similarly as for the metric and
Palatini formalism, $f(X)$ gravity can be transformed into scalar-tensor
theory \cite{Harko:2011nh, Capoz, Capozziello:2013uya, Capoz2}.  
Using the respective dynamically equivalent scalar-tensor representation, it was shown that the theory can pass the Solar System observational constraints even if the scalar field is very light. This implies the existence of a long-range scalar field, which is able to modify the cosmological and galactic dynamics, but leaves the Solar System unaffected. The absence of instabilities in perturbations was also verified and explicit models, consistent with local tests, lead to the late-time cosmic speed-up.

It was also shown that the theory can be  formulated in terms of the quantity $X=\kappa^2 T + R$, where $T$ and $R$ are the traces of the stress-energy and Ricci tensors, respectively. The variable $X$ represents the deviation with respect to the field equation trace of GR. The cosmological applications of this metric-Palatini  gravitational theory were explored, and cosmological solutions, coming from the scalar-tensor representation of $f(X)$-gravity, have been derived \cite{Capoz}. Furthermore, criteria to obtain cosmic acceleration were discussed and the field equations were analyzed as a dynamical system. Several classes of dynamical cosmological solutions, depending on the functional form of the effective scalar field potential, describing both accelerating and decelerating Universes have been explicitly obtained.  Furthermore, the cosmological perturbation equations were derived and applied to uncover the nature of the propagating scalar degree of freedom and the signatures that  these models predict in the large-scale structure. The Cauchy problem was also explored and it was shown that the initial value problem can always be well-formulated and, furthermore, can be well-posed depending on the adopted matter sources \cite{Capozziello:2013gza}. Furthermore, the possibility that wormholes are supported in the metric-Palatini gravitational theory was also explored and it was shown that  the higher-curvature derivatives  sustain these exotic spacetimes \cite{Capozziello:2012hr}. 

At the intermediate galactic scale, the possibility that the behavior of the rotational velocities of test particles gravitating around galaxies could be explained within the framework of the hybrid metric-Palatini gravitational theory was also considered \cite{Capozziello:2013yha}. It was shown  that the circular velocity can be explicitly obtained as a function of the scalar field of the equivalent scalar-tensor description. The possibility of constraining the form of the scalar field potential and the parameters of the model by using the stellar velocity dispersions was also analyzed. All the physical and geometrical quantities and the numerical parameters in hybrid metric-Palatini gravity were  expressed in terms of observable/measurable parameters, such as the circular velocity, the baryonic mass of the galaxy, the Doppler frequency shifts, and the stellar dispersion velocity, respectively. Therefore, the obtained results open the possibility of testing the Hybrid  Gravity  at galactic and extra-galactic scales by using direct astronomical observations.

In this context, the virial theorem was also generalized in the scalar-tensor representation of the  metric-Palatini gravity \cite{Capozziello:2012qt}. More specifically, taking into account the relativistic collisionless Boltzmann equation, it was shown that the supplementary geometric terms in the gravitational field equations provide an effective contribution to the gravitational potential energy. It was shown that the total virial mass is proportional to the effective mass associated with the new terms generated by the effective scalar field, and the baryonic mass. This shows that the geometric origin in the generalized virial theorem may account for the well-known virial theorem mass discrepancy in clusters of galaxies. In addition to this, astrophysical applications of the model were also considered and shown that the model predicts that the mass associated to the scalar field and its effects extend beyond the virial radius of the clusters of galaxies. According to the galaxy cluster velocity dispersion profiles,  the generalized virial theorem can be an efficient tool in testing observationally  the viability of this class of  gravity models.

However, the problem of selecting viable models cannot be posed only at a phenomenological level but should be considered also at the fundamental level. To this end, symmetries are extremely useful  to fix self-consistent  models. In particular, Noether symmetries, beside selecting conserved quantities,   are also useful to reduce dynamical systems and find out exact solutions. In cosmology, the so-called Noether Symmetry Approach revealed extremely useful to work out physically motivated models related to conservation laws \cite{Cap96}.

The Noether symmetry approach has been applied in cosmology by many authors in various contexts, such as in scalar-tensor gravity \cite{Capozziello:1994du}, higher order gravity \cite{CamL}, and in teleparallel gravity \cite{BasFT}. 
For instance, new exact solutions for cosmological models with a minimally coupled  scalar field were first  found by requiring the existence of a Noether symmetry for a Lagrangian description on a two-dimensional ``configuration space'' \cite{deRiti90}. 
Furthermore, the evolution of  two dimensional minisuperspace cosmological models, at the classical and quantum levels, was investigated and exact solutions achieved by using  the Noether Symmetry Approach considering, as  phase space variables, the FRW scale factor and the  scalar field \cite{VF12}. 
Furthermore,  the Noether  Symmetry Approach can be applied to quantum cosmology \cite{Cap96, Aslam}, phantom quintessence cosmology \cite{CapP09}, to spinor and scalar field models \cite{Kremer:2013gd}. Finally, the dynamics of homogeneous cosmologies with a scalar field  source with an arbitrary self-interaction potential was also explored in \cite{KotsakisL}. Bianchi universes and related Noether symmetries have been considered in \cite{Cap97M}. 

Indeed, the literature on the applications of  Noether symmetries in extensive, and we mention only  a few works relative to the analysis of nonminimally coupled scalar field cosmologies. For instance, in \cite{Cap93deR}, a systematic analysis of nonminimally coupled cosmologies in $(n+1)-$dimensional homogeneous and isotropic spacetimes was performed. In \cite{Souza}, nonminimally coupled scalar fields were studied in the Palatini formalism, and in \cite{Capozziello:1994du}, applying Noether's symmetries exact solutions in flat nonminimally coupled cosmological models were obtained.
In Ref. \cite{deSouza:2008az}, a cosmological model where a fermion field is nonminimally coupled with the gravitational field was analysed. By applying the Noether Symmetry Approach, cosmological solutions were found showing that  a nonminimally coupled fermion field behaves as an inflaton describing an accelerated inflationary scenario, whereas the minimally coupled fermion field describes a decelerated period being identified as dark matter.

In the higher gravity context, the application of the Noether theorem is a powerful tool to find  solutions of the field equations and select viable models \cite{Sanyal05}. In \cite{CamL} it was shown that higher-order corrections of the Einstein-Hilbert action of General Relativity can be recovered by imposing the existence of a Noether symmetry to a class of theories of gravity where the Ricci scalar $R$ and its d'Alembertian $\Box R$ were present.
It was also shown that by using the Noether symmetries in Minisuperspace Quantum Cosmology, the existence of conserved quantities provides selection rules to recover  classical behaviors in the cosmic evolution \cite{Capozziello:2013qha}.
In \cite{defelice, Vakili:2010dh}, it was shown that the existence of  Noether symmetries in  minisuperspace provides the form of the $f(R)$ function  and the   cosmic evolution of the scale factor. In \cite{dimakisT}, considering a FRW space-time and a perfect fluid described by a barotropic equation of state,  the reparametrization invariance of the resulting $f(R)$ Lagrangian was used to work in the equivalent constant potential description. In addition to this, the integrals of motion were used to analytically solve the equations of the corresponding models. Indeed, the application to modified gravity is extensive, and in addition to the above work, we refer to the Noether symmetric $f(R)$ quantum cosmology and its classical correlations \cite{VakF}, to  the Lie point and variational symmetries in minisuperspace Einstein gravity \cite{Christod}, and to the Noether Symmetry in $f(T)$ theory, where $T$ is the trace of the torsion tensor \cite{Hwei}. Scalar tensor teleparallel dark gravity has also been explored via the Noether Symmetry Approach in \cite{Kucukakca:2013mya}. The Noether gauge symmetry for $f(R)$ gravity in the Palatini formalism has been developed in  \cite{Camci}; Hamiltonian dynamics and Noether symmetries in extended gravity cosmology has been considered \cite{CapHD}.

In this paper, we will consider point symmetries in order to obtain exact
solutions of the field equations and invariant solutions for the Hybrid Gravity Wheeler-DeWitt
Equation (WDW)  in a spatially flat Friedmann-Robertson-Walker (FRW) spacetime.  In fact, the
infinitesimal generator of a point transformation, which leaves invariant the
field equations, is a Noether symmetry. This feature provides integrals of motions capable of
reducing the related dynamical system and then getting exact solutions. 
For the determination of the Noether symmetries of the classical Lagrangian, we
will apply the geometric procedure outlined in \cite{TsamAnd}, where the Noether
symmetries of the Lagrangian are connected to the collineations of the second order tensor
which is defined by the kinematic part of the Lagrangian. Therefore the Noether
symmetry is not only a criterion for the integrability of the system but 
also a geometric criterion since it allows  to select the free
functions of the theory. This approach has
been applied in \cite{TsamC01,TsamC02,PTB,BasSF,BasFT,AndFT}. Furthermore, in
order to solve exactly the WDW equation we will apply the theory of Lie
invariants that allows to determine the Lie point symmetries of the WDW equation. As shown in 
 \cite{AnIJGMP}, the Lie point
symmetries for the WDW equation (or the Klein Gordon equation) are connected to
the conformal Lie algebra of the minisuperspace which defines the Laplace operator.

The outline of this paper is the following. In Sec. \ref{FieldEquations}, we
introduce the basic definitions of the Hybrid Gravity and we show that it is
possible to write the field equations  in terms of a
nonminimally coupled scalar field theory. In Sec. \ref{NPS}, we apply the Noether Symmetry Approach  to the Hybrid Gravity Lagrangian.
The only case that admits Noether symmetries is the linear $f(\mathcal{R})$ function. Due to this feature, in Sec. \ref{ConT},  we study the integrability of the field
equations in conformal frames  to find out new solutions. In Sec. \ref{Exact}, we apply the extra
Noether symmetries and solve the field equations. Furthermore, we determine
the WDW equation for each case and apply the zero order invariants of the Lie
point symmetries to reduce the order of the partial differential equation and
to find invariant solutions. Finally, in section \ref{Conclusion}, we draw
our conclusions.

\section{Hybrid Gravity Cosmology}

\label{FieldEquations}

Let us consider the action for the hybrid metric-Palatini gravity in the following form \cite{Harko:2011nh,Capoz} :
\begin{equation}
S=\frac{1}{2\kappa^{2}}\int\mathrm{d}^{4}x\sqrt{-g}[R+f(\mathcal{R})]+S_{m},
\label{action}%
\end{equation}
where $R$ is the metric curvature scalar and $f(\mathcal{R})$ is a function
of the Palatini curvature scalar which is constructed through an independent
connection $\hat{\Gamma}$. The variation of the above action with respect to the
metric yields the gravitational field equations
\begin{equation}
\label{eq_hyb}G_{\mu\nu}+f^{\prime}(\mathcal{R})\mathcal{R}_{\mu\nu}-\frac
{1}{2}f(\mathcal{R})g_{\mu\nu}=\kappa^{2}T_{\mu\nu},
\end{equation}
where $G_{\mu\nu}$ is the Einstein tensor for the metric $g_{\mu\nu}$ (with
Lorentzian signature), while $\mathcal{R}_{\mu\nu}$ is a Ricci tensor
constructed by the conformally related metric $h_{\mu\nu}=f^{\prime
}(\mathcal{R})g_{\mu\nu}$, where the conformal factor is given by $f'(\mathcal{R})=df(\mathcal{R})/d \mathcal{R}$. The trace of Eq. (\ref{eq_hyb}) is the hybrid
structural equation, where one can algebraically express the Palatini curvature $\mathcal{R}$
 in terms of a quantity $X$ assuming that $f(\mathcal{R})$ has analytic
solutions, that is:
\begin{equation}
f^{\prime}(\mathcal{R})\mathcal{R}-2f(\mathcal{R})=\kappa^{2}T+R\equiv X.
\label{master}%
\end{equation}
The variable $X$ measures the deviation from the GR  trace equation $R=-\kappa^{2}T$, that is GR is fully recovered for $X=0$ \cite{Capoz}.

As for the pure metric and Palatini case \cite{report,olmo}, the above
action can be transformed into a scalar-tensor theory by introducing an
auxiliary field $E$ such that
\begin{equation}
S=\frac{1}{2\kappa^{2}}\int\mathrm{d}^{4}x\sqrt{-g}[R+f(E)+f^{\prime
}(E)(\mathcal{R}-E)].
\end{equation}
The field $E$ is dynamically equivalent to the Palatini scalar $\mathcal{R}$ if
$f^{\prime\prime}(\mathcal{R})\neq0$. Defining$\;$%
\begin{equation}
\phi\equiv f^{\prime}(E), \qquad   V(\phi)=Ef^{\prime}(E)-f(E), \label{lagr}%
\end{equation}
the action becomes
\begin{equation}
S=\frac{1}{2\kappa^{2}}\int\mathrm{d}^{4}x\sqrt{-g}[R+\phi\mathcal{R}%
-V(\phi)].
\end{equation}
Using the relation between $R$ and $\mathcal{R}$, given by
\begin{equation} \label{eq:conformal_R}
\mathcal{R}=R+\frac{3}{2\phi^2}\partial_\mu \phi \partial^\mu \phi-\frac{3}{\phi}\Box \phi,
\end{equation}
(see \cite{Capoz} for details) one finally obtains the standard scalar-tensor form
\begin{equation}
S=\frac{1}{2\kappa^{2}}\int\mathrm{d}^{4}x\sqrt{-g}\left[(1+\phi)R+\frac{3}{2\phi
}\partial^{\mu}\phi\partial_{\mu}\phi-V(\phi)\right]. \label{lagr0}%
\end{equation}

It is  worth noticing that Eq. (\ref{lagr}) is a Clairaut differential  equation \cite{Claraut}, that is,
\begin{equation}
Ef^{\prime}(E)-f(E)=V\left(  f^{\prime}\left(  E\right)  \right)  .
\label{fe.04}%
\end{equation}
It admits a general linear solution%
\begin{equation}
f\left(  E\right)  =cE-V\left(  c\right)  \label{fe.05}%
\end{equation}
for arbitrary $V\left(  \phi\right)  $ and a singular solution followed from
the equation
\begin{equation}
\frac{\partial V\left(  f^{\prime}\left(  E\right)  \right)  }{\partial
f^{\prime}}-E=0. \label{fe.06}%
\end{equation}

Let us consider the FRW spatially flat metric
\begin{equation}
ds^{2}=-dt^{2}+a^{2}\left(  t\right)  \left(  dx^{2}+dy^{2}+dz^{2}\right)  .
\label{FRW}%
\end{equation}
Therefore, from action (\ref{lagr0}) one  deduces the pointlike Lagrangian
\begin{equation}
\mathcal{L}=6a\dot{a}^{2}(1+\phi)+6a^{2}\dot{a}\dot{\phi}+\frac{3}{2\phi}%
a^{3}\dot{\phi}^{2}+a^{3}V(\phi), \label{lagr1}%
\end{equation}
from which  the following field equations can be deduced:
\begin{equation}
\ddot{a}+\frac{1-\phi}{2a}\dot{a}^{2}-\frac{1}{2}\dot{a}\dot{\phi}-\frac
{a}{3\phi}\dot{\phi}^{2}-\frac{1}{12}a\left(  3V-2\phi V_{,\phi}\right)  =0 ,
\label{fe.02}%
\end{equation}
and
\begin{eqnarray}
\ddot{\phi}+\frac{\phi\left(  \phi+1\right)  }{a^{2}}\dot{a}^{2}+\frac{\phi
+3}{a}\dot{a}\dot{\phi}+\frac{\phi-2}{4\phi}\dot{\phi}^{2}
    \nonumber  \\
+\frac{\phi}%
{6}\left(  3V\left(  \phi\right)  -2\left(  \phi+1\right)  V_{,\phi}\right)
=0  \,. \label{fe.03}%
\end{eqnarray}
Note that Eq. (\ref{fe.03}) is the Klein-Gordon equation for the scalar field $\phi$.
 The energy condition is given by
\begin{equation}
6a\dot{a}^{2}(1+\phi)+6a^{2}\dot{a}\dot{\phi}+\frac{3}{2\phi}a^{3}\dot{\phi
}^{2}-a^{3}V(\phi)=0. \label{fe.01}%
\end{equation}

Equations (\ref{fe.02}) and (\ref{fe.01}) can be written in the form of  modified Friedmann
equations for the scale factor $a(t)$, that is 
\begin{align*}
3H^{2}  &  = \kappa^{2}\rho_{\rm eff} , \\
\left(  2\dot{H}+3H^{2}\right)   &  = -\kappa^{2}p_{\rm eff},
\end{align*}
where $H=\dot{a}/a$ is a Hubble parameter and $(\rho_{\rm eff}$ and $p_{\rm eff})$
are the total effective energy density  and pressure  given by
\begin{align}
\rho_{\rm eff}  &  =\frac{1}{\kappa^{2}}\frac{2\phi V(\phi)-12H\phi\dot{\phi
}-3\dot{\phi}^{2}}{6\phi(1+\phi)}\label{fe.03a}\\
p_{\rm eff}  &  =\frac{1}{\kappa^{2}}\frac{2\phi^{2}V_{,\phi}-3\phi V(\phi
)-6\phi^{2}H^{2}-6H\phi\dot{\phi}-4\dot{\phi}^{2}}{6\phi}  ,  \label{fe.03b}%
\end{align}
respectively.  


\section{The Noether Symmetry Approach }
\label{NPS}

\subsection{General considerations}

Let us search for   Noether point symmetries in order to determine the form of the potential $V(\phi) $ 
in Eq. (\ref{lagr1}). Besides,  the method will allow also to   derive the singular
solution for the  Clairaut Eq. (\ref{fe.04}). Furthermore, we will apply the Noether conservation laws for the
reduction of the dynamical system. Finally,  we will  search for   
Lie point symmetries in the WDW equation in order to find invariant solutions.

Before proceeding further, we briefly review  the basic definitions concerning
Noether symmetries for systems of second order ordinary differential equations
(ODEs) of the form
\begin{equation}
\ddot{x}^{\alpha}=\omega^{\alpha}\left(  t,x^{\beta},\dot{x}^{\beta}\right)
.\label{Lie.0}%
\end{equation}
Let the system of ODEs (\ref{Lie.0}) result from a first order Lagrangian
${\cal L}={\cal L}\left(  t,x^{\beta},\dot{x}^{\beta}\right)  $. Then the vector field
\[
X=\xi\left(  t,x^{\beta}\right)  \partial_{t}+\eta^{\alpha}\left(  t,x^{\beta
}\right)  \partial_{\alpha}  ,
\]
in the augmented space $\{t,x^{i}\}$ is a generator of a Noether point
symmetry for the ODEs system (\ref{Lie.0}),  if the additional condition 
\begin{equation}
X^{\left[  1\right]  }{\cal L}+{\cal L}\frac{d\xi}{dt}=\frac{dg}{dt} , \label{Lie.5}
\end{equation}
is
satisfied~\cite{StephaniB}, where $g=g\left(  t,x^{\beta}\right) $ is the gauge function and $X^{\left[
1\right]  }$ is the first prolongation of $X$, i.e.,
\[
X^{\left[  1\right]  }=X+\left(  \frac{d\eta^{\beta}}{dt}-\dot{x}^{\beta}%
\frac{d\xi}{dt}\right)  \partial_{\dot{x}^{\beta}}.
\]
For every Noether point symmetry there exists a first integral (a Noether
integral) of the system of equations (\ref{Lie.0}) given by the formula:%
\begin{equation}
I=\xi E_{H}-\frac{\partial {\cal  L}}{\partial\dot{x}^{\alpha}}\eta^{\alpha}+g , 
\label{Lie.6}
\end{equation}
where $E_{H}$ is the Hamiltonian of ${\cal L}$.

The vector field $X$ for the Lagrangian (\ref{lagr1}) is
\begin{equation}
X=\xi\left(  t,a,\phi\right)  \partial_{t}+\eta_{a}\left(  t,a,\phi\right)
\partial_{a}+\eta_{\phi}\left(  t,a,\phi\right)  \partial_{\phi} , \label{Lie.3}
\end{equation}
and the first prolongation is given by
\begin{equation}
X^{\left[  1\right]  }=\xi\partial_{t}+\eta_{a}\partial_{a}+\eta_{\phi
}\partial_{\phi}+\left(  \dot{\eta}_{a}-\dot{a}\dot{\xi}\right)
\partial_{\dot{a}}+\left(  \dot{\eta}_{\phi}-\dot{\phi}\dot{\xi}\right)
\partial_{\dot{\phi}}. \label{Lie.7}%
\end{equation}
These considerations can be immediately  applied to the Hybrid Gravity.

\subsection{Searching for Noether Symmetries in Hybrid  Gravity}

Lagrangian (\ref{lagr1}) is in the standard form ${\cal L}=T-V_{\rm eff}$, where
$T=\frac{1}{2}g_{\mu\nu}\dot{x}^{\mu}\dot{x}^{\nu}$ is the kinetic energy with a
``kinetic'' metric
\begin{equation}
ds_{\left(  2\right)  }^{2}=12a\left(  1+\phi\right)  da^{2}+12a^{2}%
dad\phi+\frac{3}{\phi}a^{3}d\phi^{2} , \label{HG.02}%
\end{equation}
and effective potential
\begin{equation}
V_{\rm eff}=-a^{3}V\left(  \phi\right)  . \label{HG.03}%
\end{equation}
Therefore, in order to search for special forms of the potential $V\left(
\phi\right) $, where the Lagrangian admits Noether point symmetries, we will
apply the geometric approach developed in  \cite{TsamAnd}.

Since the Lagrangian is time-independent, it admits the Noether symmetry
$\partial_{t}$ with  the Hamiltonian as a conservation law, that is
\begin{equation}
E_{H}=6a\left(  1+\phi\right)  \dot{a}^{2}+6a^{2}\dot{a}\dot{\phi}+\frac
{3}{2\phi}a^{3}\dot{\phi}^{2}-a^{3}V\left(  \phi\right)\,.
\end{equation}
Due to the Einstein  equation $G_{0}^{0}=0$, which is a constraint, we have  $E_{H}=0$ in vacuum.

Following the results of \cite{TsamAnd},  the Lagrangian (\ref{lagr1})
admits an extra Noether symmetry in the case of a constant potential 
$V(\phi)  =V_{0}$. Specifically,  for the constant potential, the Noether
symmetry is given by
\[
X_{1}=\frac{\sqrt{\phi}}{a}\partial_{\phi},
\]
and the corresponding Noether integral has the form%
\[
I_{1}=3\frac{a}{\sqrt{\phi}}\left(  2\phi\dot{a}+a\dot{\phi}\right)  .
\]

Under the coordinate transformation%
\[
a=u^{\frac{2}{3}}, \qquad \phi=v^{2}u^{-\frac{4}{3}},
\]
the Lagrangian of the field equations become%
\[
{\cal L}\left(  u,v,\dot{u},\dot{v}\right)  =\frac{8}{3}\dot{u}^{2}+6u^{\frac{2}{3}%
}\dot{v}^{2}+V_{0}u^{2} .
\]
The field equations are given by
\begin{equation}
\frac{8}{3}\dot{u}^{2}+6u^{\frac{2}{3}}\dot{v}^{2}-V_{0}u^{2}=0 , \label{HG.04a}
\end{equation}
\begin{align}
\ddot{u}-\frac{3}{4}u^{-\frac{1}{3}}\dot{v}^{2}-\frac{3}{8}V_{0}u  &
=0, \label{HG.05a}\\
\ddot{v}+\frac{2}{3u}\dot{u}\dot{v}  &  =0,
\end{align}
respectively. The extra Noether integral in the $\{u,v\}$ variables can be written as
$\bar{I}_{1}=u^{\frac{2}{3}}\dot{v}$ (where $\bar{I}_{1}/I_{1}={\rm const}$) so one
has $\dot{v}=\bar{I}_{1}u^{-\frac{2}{3}}$. The general solution of the above system is 
\begin{equation}
\int\frac{du}{\sqrt{\frac{3}{8}V_{0}u^{2}-\frac{9}{4}\bar{I}_{1}^{2}%
u^{-\frac{2}{3}}}}=\int dt. \label{HG.04ab}%
\end{equation}
Furthermore, for the Hubble function $H=\dot{a}/a$, we have
\begin{equation}
\frac{H^{2}}{H_{0}^{2}}=\left(  \Omega_{\Lambda}+\Omega_{r}a^{-4}\right),
\label{HG.05aa}%
\end{equation}
where 
\begin{equation}
\Omega_{\Lambda}=\frac{1}{6}\frac{V_{0}}{H_{0}^{2}}, \qquad {\rm and} \qquad
\Omega
_{r}=-\frac{\bar{I}_{1}^{2}}{H_{0}^{2}}\,,
\end{equation}
clearly indicate the density parameters for the cosmological constant and radiation, respectively.
Note that in order to have a physical solution, it has to be  $\bar{I}_{1}\in \mathbb{C}
$ and  $\operatorname{Re}\left(  \bar{I}_{1}\right)  =0$.

The Hubble function (\ref{HG.05aa}) corresponds to the model with \ a
cosmological constant and a radiation fluid. However ,if we  introduce dust
in our model, $\rho_{D}=\rho_{m0}a^{-3}$,  Eq. (\ref{HG.04a}) becomes
\[
\frac{8}{3}\dot{u}^{2}+6u^{\frac{2}{3}}\dot{v}^{2}-V_{0}u^{2}=\rho_{m0}.
\]
Therefore, the analytical solution  takes the form
\[
\int\frac{du}{\sqrt{\frac{3}{8}V_{0}u^{2}+\frac{3}{8}\rho_{m0}-\frac{9}{4}%
\bar{I}_{1}^{2}u^{-\frac{2}{3}}}}=\int dt ,
\]
and the Hubble function is
\[
\frac{H^{2}}{H_{0}^{2}}=\left(  \Omega_{\Lambda}+\Omega_{m}a^{-3}+\Omega
_{r}a^{-4}\right)  ,
\]
where now ${\displaystyle \Omega_{m}=\frac{\rho_{m0}}{6H_{0}^{2}}}$. Thus, the Hybrid Gravity introduces a further ``radiation'' term.

Since the linear case is trivial, in the next section, we will perform a
conformal transformation for the Lagrangian (\ref{lagr1}) in order to apply the
results of \cite{TsamC01,TsamC02}. We will consider two separate cases with
respect to a lapse function $N$: $(i)$ the case where the lapse is a function of the
scale factor, $d\tau=N\left(  a\right)  dt$, and $(ii)$ the case where it is a function of the
scalar field, i.e. $d\tau=N\left(  \phi\right)  dt$. For the latter case we will
show that Hybrid Gravity is conformally related to a Brans-Dicke-like scalar field  theory.

\section{Conformal transformations}
\label{ConT}

The dynamical system described by the Lagrangian (\ref{lagr1}) is conformally
invariant, with $E_{H}=0$. Hence, we can apply conformal transformations to
the Lagrangian (\ref{lagr1}) in order to use the results in
\cite{TsamC01,TsamC02,AnIJGMP} and determine new solutions in conformal
frames. However, since the minisuperspace described by the metric (\ref{HG.02}) is two-dimensional, it
admits an infinite conformal algebra, so that in order to simplify the problem we have to
provide some ansantz. As a first one, we will consider conformal
transformation of the form $d\tau=N\left(  a\right)  dt$, so that the
spacetime metric (\ref{FRW}) has the form
\begin{equation}
ds^{2}=-N^{-2}\left(  a\left(  \tau\right)  \right)  d\tau^{2}+a^{2}\left(
\tau\right)  \left(  dx^{2}+dy^{2}+dz^{2}\right)  . \label{FRW1}%
\end{equation}
Then we will study the case of conformal transformations of the form
$d\tau=N\left(  \phi\right)  dt$.

\subsection{Noether Symmetries for the conformal Lagrangian}

The Lagrangian of the field equations for the conformal FRW spacetime
(\ref{FRW1}) is given by
\begin{eqnarray}
&& {\cal L}\left(  a,\phi,a^{\prime},\phi^{\prime}\right)  =  \frac{a^{3}V\left(  \phi\right)
}{N\left(  a\right)  } 
\nonumber \\
&&+N\left(  a\right)  \left[
6a\left(  1+\phi\right)  a^{\prime2}+6a^{2}a^{\prime}\phi^{\prime}
     +\frac{3}{2\phi}a^{3}\phi^{\prime2}\right] , \label{FRW.02}%
\end{eqnarray}
where the prime denotes $d/d\tau$ (it should not be confused with the
conformal time that requires a special choice of the lapse
function $N(a)$). The conformal kinetic metric and the related Ricci scalar are given by
\begin{equation}
d\bar{s}_{\left(  2\right)  }^{2}=N\left(  a\right)  \left(  12a\left(
1+\phi\right)  da^{2}+12a^{2}dad\phi+\frac{3}{\phi}a^{3}d\phi^{2}\right) ,
\label{FRW.03}%
\end{equation}
and
\[
R_{\left(  2\right)  }=-\frac{a^{2}NN_{,aa}-a^{2}N_{,a}^{2}-N^{2}}%
{12a^{3}N^{3}},
\]
respectively. Since the kinetic metric (\ref{FRW.03}) is two-dimensional, the space is
an Einstein space. If the Ricci scalar is constant, the Einstein space  has a constant curvature. In order to reduce the problem to the dynamics of
Newtonian physics \cite{TsamC02}, we consider $R_{\left(  2\right)  }=0$, so that
\begin{equation}
N\left(  a\right)  =a^{-1}e^{N_{0}a}. \label{FRW.04}%
\end{equation}
Therefore, by applying the geometric approach in  \cite{TsamAnd},  one gets that
the Lagrangian (\ref{FRW.02}), with the solution (\ref{FRW.04}), admits  extra Noether
symmetries. The first one is of the form
\begin{equation}
X_{1}=-\frac{1}{2}\partial_{a}+\frac{\phi+V_{1}\sqrt{\phi}}{a}\partial_{\phi},
\label{FRW.05}%
\end{equation}
with the corresponding conservation law
\begin{equation}
I_{X_{1}}=6\left(  V_{1}\sqrt{\phi}-1\right)  \dot{a}+3\frac{a}{\sqrt{\phi}%
}V_{1}\dot{\phi} , \label{FRW.06}%
\end{equation}
for the potential
\begin{equation}
V\left(  \phi\right)  =V_{0}\left(  \sqrt{\phi}+V_{1}\right)  ^{4}.
\label{FRW.07}%
\end{equation}

The second symmetry vector and corresponding conservation law are given by
\begin{equation}
X_{2}=2\tau\partial_{\tau}+a\left(  \sqrt{\phi}V_{1}+1\right)  \partial
_{a}-2V_{1}\sqrt{\phi}\left(  \phi+1\right)  \partial_{\phi}  , \label{FRW.08}%
\end{equation}
and
\begin{equation}
I_{X_{2}}=12a\left(  1+\phi\right)  \dot{a}+6a^{2}\left(  1-\frac{V_{1}}%
{\sqrt{\phi}}\right)  \dot{\phi},  \label{FRW.09}%
\end{equation}
respectively, with the potential given by
\begin{equation}
V\left(  \phi\right)  =V_{0}\left(  1+\phi\right)  ^{2}\exp\left(  \frac
{6}{V_{1}}\arctan\sqrt{\phi}\right)  . \label{FRW.10}%
\end{equation}
We have chosen $N_{0}=0$ for both cases. In the case of $N_{0}\neq0$, one finds
that the Lagrangian (\ref{FRW.02}) admits extra Noether symmetries only in the
case of the trivial potential $V\left(  \phi\right)  =0$.

Let us stress that the Noether Integrals (for both cases), the
Hamiltonian$~E_{H}$ and $I_{X}$ are independent geometrical objects. 
The relation $\left\{
E_{H},I_{X}\right\}  =0$ holds, therefore the dynamical systems are
Liouville integrable.

Furthermore, the Clairaut Eq. (\ref{lagr}) for the potential
(\ref{FRW.07}) is given by
\begin{equation}
Ef^{\prime}(E)-f(E)=V_{0}\left(  \sqrt{f^{\prime}\left(  E\right)  }%
+V_{1}\right)  ^{4}.
\end{equation}
Thus, Eq. (\ref{fe.06}) becomes%
\begin{equation}
\frac{2V_{0}}{\sqrt{f^{\prime}\left(  E\right)  }}\left(  \sqrt{f^{\prime
}\left(  E\right)  }+V_{1}\right)  ^{3}+E=0 ,
\end{equation}
and hence, by setting $y=\sqrt{f^{\prime}\left(  E\right)  }$, one has the
polynomial equations%
\begin{equation}
\left(  y+V_{1}\right)  ^{3}-\frac{E}{2V_{0}}y=0 . \label{HG.051g1}%
\end{equation}
A real solution of Eq. (\ref{HG.051g1}) is
\begin{equation}
\int df=\int\left(  EF\left(  E\right)  +\frac{1}{6V_{0}F\left(  E\right)
}-V_{1}\right)  ^{2}dE ,
\end{equation}
where
\begin{equation}
F^{3}\left(  E\right)  =6EV_{0}^{2}\left(  \sqrt{81V_{1}^{2}-\frac{6E}{V_{0}}%
}-9V_{1}\right)  .
\end{equation}

Let us simplify the equation. For instance, considering $V_{1}=0$, the singular solution of
the Clairaut equation yields
\begin{equation}
f\left(  E\right)  =\frac{E^{2}}{4V_{0}}\text{.} \label{GR_sol}%
\end{equation}
It should be noticed that if one substitutes the solution (\ref{GR_sol}) into
the hybrid master equation (\ref{master}) one finds that the variable $X=0$,
that is, the solution is the GR case.

Similarly, we can proceed for the potential given by (\ref{FRW.10})
\begin{eqnarray}
Ef^{\prime}(E)-f(E)=V_{0}\left[1+f^{\prime}\left(  E\right)  \right]
^{2}\times
    \nonumber \\
\times \exp\left(  \frac{6}{V_{1}}\arctan\sqrt{f^{\prime}\left(  E\right)
}\right)\,,
\end{eqnarray}
so that the singular solution follows from the equation%
\begin{eqnarray}
&& \left[ 1+f^{\prime}\left(  E\right)  \right]  \left(  2+\frac{3}{V_{1}%
\sqrt{f^{\prime}\left(  E\right)  }}\right)  \times
   \nonumber \\
&&\times \exp\left(  \frac{6}{V_{1}%
}\arctan\sqrt{f^{\prime}\left(  E\right)  }\right)  +E=0\,.
\end{eqnarray}

\subsection{Hybrid Gravity as a Brans-Dicke-like scalar field}

Let us apply now  the conformal transformation $\bar{g}%
_{ij}=N\left(  \phi\right)  ^{-2}g_{ij}$ in the FRW spacetime (\ref{FRW}).
Under this transformation, the action of Hybrid Gravity (\ref{lagr0}) becomes%
\begin{equation}
S=\frac{1}{2\kappa^{2}}\int d^{4}x\sqrt{-\bar{g}}[(1+\phi)\bar{R}+\frac
{3}{2\phi}\bar{g}^{ij}\phi_{,i}\phi_{,j}-V(\phi)], \label{HG.01}%
\end{equation}
where the conformal Ricci scalar is given by
\[
\bar{R}=N^{-2}R-6N^{-3}g^{ij}N_{;ij}.
\]
Substituting it into the action (\ref{HG.01}) one finds%
\begin{eqnarray}
S&=&\frac{1}{2\kappa^{2}}\int d^{4}x\sqrt{-g}[(1+\phi)N^{2}R-6(1+\phi
)Ng^{ij}N_{;ij}
    \nonumber  \\
&&+\frac{3}{2\phi}N^{2}g^{ij}\phi_{,i}\phi_{,j}-N^{4}V(\phi)] .
\label{HG.01a}%
\end{eqnarray}
If we take into account the following lapse function%
\begin{equation}
N\left(  \phi\right)  =\sqrt{\frac{F\left(  \phi\right)  }{1+\phi}},
\end{equation}
we have
\begin{equation}
N_{;i}=\frac{1}{2}\sqrt{\frac{1+\phi}{F\left(  \phi\right)  }}\left(
\frac{F_{,\phi}}{1+\phi}-\frac{F}{\left(  1+\phi\right)  ^{2}}\right)
\phi_{;i}.
\end{equation}
Substituting  the results into  the various terms of Eq. (\ref{HG.01a}), we get the following relations
\begin{equation}
\int d^{4}x\sqrt{-g}\left[  \left(  1+\phi\right)  N^{2}R\right]  =\int
d^{4}x\left[  F\left(  \phi\right)  R\right] ,
\end{equation}
\begin{eqnarray}
\int d^{4}x\sqrt{-g}\frac{3}{2\phi}N^{2}g^{ij}\phi_{,i}\phi_{,j}&=&
    \nonumber \\
&& \hspace{-2.5cm} \int d^{4}x\sqrt{-g}\frac{3F\left(  \phi\right)  }{2\left(  1+\phi\right)  \phi
}g^{ij}\phi_{,i}\phi_{,j} ,
\end{eqnarray}
and a lengthy, but straightforward calculation leads to
\begin{eqnarray}
&&\int d^{4}x\sqrt{-g}6(1+\phi)Ng^{ij}N_{;ij}   = 
    \nonumber \\
&& \hspace{-0.5cm} - \int d^{4}x\sqrt{-g}\left[  \frac{3}{2}\frac{F\left(  \phi\right)
-\left(  1+\phi\right)  ^{2}F_{,\phi}^{2}}{\left(  1+\phi\right)  F\left(
\phi\right)  }g^{ij}\phi_{;i}\phi_{;j}\right] .
\end{eqnarray}

Thus, using the above relations, the action (\ref{HG.01a}) takes the form
\begin{eqnarray}
S=\int d^{4}x\sqrt{-g}\Bigg[  F\left(  \phi\right)  R+\frac{3}{2}\Bigg(
\frac{F\left(  \phi\right)  -\left(  1+\phi\right)  ^{2}F_{,\phi}^{2}}{\left(
1+\phi\right)  F\left(  \phi\right)  }
    \nonumber \\
+\frac{3F\left(  \phi\right)  }{2\left(
1+\phi\right)  \phi}\Bigg)  g^{ij}\phi_{,i}\phi_{,j}-\frac{F^{2}\left(
\phi\right)  }{\left(  1+\phi\right)  ^{2}}V(\phi)\Bigg] . \label{HG.03a}%
\end{eqnarray}

Moreover,  considering the new scalar field~$\Phi$ (i.e. a coordinate
transformation) as follows%
\[
d\Phi=\sqrt{3\left[  \frac{F\left(  \phi\right)  -\left(  1+\phi\right)
^{2}F_{,\phi}^{2}}{\left(  1+\phi\right)  F\left(  \phi\right)  }%
+\frac{3F\left(  \phi\right)  }{2\left(  1+\phi\right)  \phi}\right]  } \; d\phi,
\]
the action becomes%
\begin{equation}
S=\int d^{4}x\sqrt{-g}\left[  F\left(  \Phi\right)  R+\frac{1}{2}g^{ij}%
\Phi_{,i}\Phi_{,j}-\bar{V}\left(  \Phi\right)  \right],  \label{HG.04}%
\end{equation}
where 
\begin{equation}
\bar{V}\left(  \Phi\right)  =\frac{F^{2}\left(  \Phi\right)  }{\left(
1+\Phi\right)  ^{2}}V(\Phi).
\end{equation}
The Noether symmetry classification for the
Lagrangian in the action (\ref{HG.04}) has been completely achieved in
\cite{TsamC02} and previously in \cite{Cap96}. When $F\left(  \Phi\right)  =F_{0}\Phi^{2}$~we have a
Brans-Dicke-like scalar field with a potential. However, for  $F_{0}=-1/12$ the Brans-Dicke
scalar field gives $\omega_{0}=-3/2$ and the Lagrangian of the field
equations is singular \cite{Capozziello:1994du}. In that case the theory is
equivalent to the Palatini $f\left(  \mathcal{R}\right)  ~$ \cite{olmo}.
Furthermore, when $F\left(  \phi\right)  =F_{0}$, i.e., $N\left(  \phi\right)
=\sqrt{F_{0}/(1+\phi)}$, the action (\ref{HG.04}) describes a minimally
coupled scalar field and the results in \cite{BasSF} can be applied.

One can apply a conformal transformation of the form $\bar{g}^{\mu\nu}%
=N^{-2}\left(  a,\phi\right)  g^{\mu\nu}$ in order to obtain exact solutions.
It is worth noticing  that the WDW equation, coming from the Hamiltonian of the theory,  is conformally invariant
and this  means that the solutions that we find remain invariant
under  conformal transformations. The same feature holds for  classical solutions.
However, it is not always possible to write  exact solutions in a close form in any conformal frame.

\section{Exact  and invariant solutions}
\label{Exact}

In this section, we determine the exact solution of the field equations
and for the models with potentials (\ref{FRW.07}) and (\ref{FRW.10}).

\subsection{The case of the potential $V\left(  \phi\right)  =V_{0}\left(  \sqrt{\phi}%
+V_{1}\right)  ^{4}$}

\subsubsection{\textbf{Lagrangian, Hamiltonian, and field equations}}

Let us apply the coordinate transformation
\begin{equation}
a=Cv+u,~\phi=\left(  \frac{v}{Cv+u}-V_{1}\right)  ^{2},
\end{equation}
where $C=V_{1}/(1+V_{1}^{2})$. In the new coordinates, the Lagrangian (\ref{FRW.02})  becomes
\begin{equation}
{\cal L}\left(  u,v,u^{\prime},v^{\prime}\right)  =6\left(  V_{1}^{2}+1\right)
u^{\prime2}+\frac{6}{\left(  V_{1}^{2}+1\right)  }v^{\prime2}+V_{0}v^{4}.
\label{An.01}%
\end{equation}
Performing a second coordinate transformation
\begin{eqnarray}
x&=&\sqrt{12\left(  V_{1}^{2}+1\right)  }u,
    \\    
y&=&\sqrt{\frac{12}{\left(  V_{1}%
^{2}+1\right)  }}v , \label{An.02}%
\end{eqnarray}
the Lagrangian (\ref{An.01}) is
\begin{equation}
{\cal L}\left(  x, y, x^{\prime}, y^{\prime}\right)  =\frac
{1}{2}x^{\prime2}+\frac{1}{2}y^{\prime2}+\bar{V}_{0}y^{4}  ,\label{An.03}%
\end{equation}
where $\bar{V}_{0}=\frac{V_{0}}{144}\left(  V_{1}^{2}+1\right)  ^{2}$. The
Hamiltonian of the field equations is given by
\begin{equation}
\tilde{H}=\frac{1}{2}p_{x}^{\prime2}+\frac{1}{2}p_{y}^{\prime2}-\bar{V}%
_{0}y^{4}, \label{An.04}%
\end{equation}
where $p_{x},p_{y}$ are the momenta. The field equations are the Hamilton
equations%
\begin{equation}
x^{\prime}=p_{x},  \qquad   y^{\prime}=p_{y}%
\end{equation}
\begin{equation}
p_{x}^{\prime}=0, \qquad p_{y}^{\prime}=4\bar{V}_{0}y^{3}  ,
\end{equation}
and the Hamiltonian constraint is $\tilde{H}=0$. Furthermore, the Hamilton-Jacobi
equation for the Hamiltonian (\ref{An.04}) provides the following action
\begin{equation}
S=c_{1}x+\int\sqrt{2\bar{V}_{0}y^{4}-c_{1}^{2}}+S_{0} ,
\end{equation}
then the field equations reduces to%
\begin{equation}
x^{\prime}=c_{1},  \qquad   y^{\prime}=\varepsilon\sqrt{2\bar{V}_{0}y^{4}-c_{1}^{2}}  .
\label{An.04as}%
\end{equation}
Thus, the exact solutions are given by
\begin{equation}
x\left(  \tau\right)  =x_{1}\tau+x_{2} \label{An.04a}%
\end{equation}
and
\begin{equation}
\int\frac{dy}{\sqrt{2\bar{V}_{0}y^{4}-c_{1}^{2}}}=\varepsilon\left(  \tau
-\tau_{0}\right)  , \label{An.05}%
\end{equation}
respectively,
where $\varepsilon=\pm1.$ In the simplest case where $V_{1}=0$,  we have the solution $a\left(  \tau\right)  =a_{0}\tau$, but from the
condition $dt=a\left(  \tau\right)  d\tau$,  we find $\tau=\sqrt{t}$, that is
$a\left(  t\right)  =a_{0}\sqrt{t}$ which is the radiation solution.

In the case where $c_{1}=0$ and $V_{1}\neq0$ from Eqs. (\ref{An.04a}) and
(\ref{An.05}), we have that
\begin{equation}
y\left(  \tau\right)  =-\varepsilon\frac{1}{\sqrt{2V_{0}}}\frac{1}{\left(
\tau-\tau_{0}\right)  }.
\end{equation}
Hence the scale factor assumes the following form
\begin{equation}
a\left(  \tau\right)  =a_{0}\left(  \tau-\tau_{0}\right)  +a_{1}-a_{2}\frac
{1}{\tau-\tau_{0}}\,. \label{solf}%
\end{equation}
From this result,  we have
\begin{equation}
\tau-\tau_{0}=\frac{1}{2a_{0}}\left(  a-a_{1}+\varepsilon\sqrt{a^{2}%
-2aa_{1}+a_{1}^{2}+4a_{0}a_{2}}\right) ,
\end{equation}
and for the Hubble function\footnote{Recall that $H=\frac{1}{a}\frac{da}%
{dt}=\frac{1}{a^{2}}\frac{da}{d\tau}$} $H\left(  \tau\right)  = a'/a^{2}$, we have
\begin{eqnarray}
H\left(  a\right)  &=& a_{0}a^{-2}+4a_{0}^{2}a_{2}\Bigg(  a^{3}-a_{1}%
a^{2}
    \nonumber  \\
&& +\varepsilon a^{2}\sqrt{\left(  a-a_{1}\right)  ^{2}+4a_{0}a_{2}} \; \Bigg)
^{-2}.
\end{eqnarray}
Therefore, in order to have a real solution, the condition
\begin{equation}
\left(  a-a_{1}\right)  ^{2}+4a_{0}a_{2}\geq0, \qquad  a\in \mathbb{R},
\end{equation}
must hold. This means  that $a_{0}a_{2}\geq0$. Hence if $a_{2}=0$,
i.e., $V_{1}=0,$ we have again  the radiation solution.

However, in the case where $a_{0}=0$,  from Eq. (\ref{solf}), we have that $\left(
\tau-\tau_{0}\right)  = a_{2}/(a_{1}-a)$ and for the Hubble function%
\begin{equation}
H\left(  a\right)  =a_{2}^{-1}\left(  a_{1}a^{-1}-1\right)  ^{2}.
\label{Hg.00}%
\end{equation}

Assuming at the present time $\left(  a=1\right) $,  we have that $H^{2}\left(
a=1\right)  =H_{0}$ hence, from Eq. (\ref{Hg.00}), we deduce $a_{2}^{-1}%
=H_{0}/\left(  \left\vert a_{1}\right\vert +1\right)  ^{2}$. Finally,
the Hubble function can be written in the following form%
\begin{equation}
\frac{H^{2}\left(  a\right)  }{H_{0}^{2}}=\Omega_{r}a^{-4}+\Omega_{m}%
a^{-3}+\Omega_{k}a^{-2}+\Omega_{f}a^{-1}+\Omega_{\Lambda} , \label{Hg.01}%
\end{equation}
where
\begin{eqnarray}
\Omega_{f}=\frac{\left\vert 4a_{1}\right\vert }{\left(  \left\vert
a_{1}\right\vert +1\right)  ^{4}}, && \qquad \Omega_{\Lambda}=\frac{1}{\left(
\left\vert a_{1}\right\vert +1\right)  ^{4}} , \label{Hg.02a}%
    \\
\Omega_{r}=\frac{\left\vert a_{1}\right\vert ^{4}}{\left(  \left\vert
a_{1}\right\vert +1\right)  ^{4}}, && \qquad \Omega_{m}=\frac{\left\vert 4a_{1}%
\right\vert ^{3}}{\left(  \left\vert a_{1}\right\vert +1\right)  ^{4}%
} ,  
    \\
 \Omega_{k}&=&\frac{\left\vert 6a_{1}\right\vert ^{2}}{\left(  \left\vert
a_{1}\right\vert +1\right)  ^{4}}\,,\label{Hg.02}%
\end{eqnarray}
where the meaning of the symbols is straightforward. 
That means that each power term of $\sqrt{\phi}$ of the power law potential
(\ref{FRW.07}), $V\left(  \phi\right)  =V_{0}\left(  \sqrt{\phi}+V_{1}\right)
^{4},$ introduces into the Hubble function a power term of the scale factor. The
corresponding fluids are: radiation, dust, curvature-like fluid,  a dark energy fluid
with equation of state $p_{f}=-\frac{2}{3}\rho_{f}~$and a cosmological constant.
We would like to note that the curvature-like fluid follows from the hybrid
gravity and not from the geometry of the spacetime, since we have considered a
spatially flat FRW spacetime. Moreover, for large redshifts $z$ the Hubble
function (\ref{Hg.01}) behaves like the radiation solution.

Furthermore, from the conformal transformation $dt=a\left(  \tau\right)
d\tau$, we have that
\begin{equation}
\tau-\tau_{0}=\exp\left[  a_{1}a_{2}^{-1}\tau_{0}-a_{2}^{-1}t-W\left(
w\left(  t\right)  \right)  \right]  =\left(  X\left(  t\right)  \right)
^{-1} \label{Hg.03}%
\end{equation}
where $w\left(  t\right)  =-a_{1}a_{2}^{-1}\exp\left[  a_{2}^{-1}\left(
a_{1}\tau_{0}-t\right)  \right]  $ and $W\left(  t\right)  $ is the Lambert
$W$-function \cite{lambert} by which we can write the exact solution for the scale
factor $a\left(  t\right)  $. By replacing Eq. (\ref{Hg.03}) in Eq. (\ref{solf}), we
find that the scale factor is expressed in terms of the proper time $t$ as follows
\begin{equation}
a^{2}\left(  t\right)  =\left[  a_{2}X\left(  t\right)  -a_{1}\right]  ^{2} .
\label{Sfhg.01}%
\end{equation}
However, from the singularity constraint $a\left(  t\rightarrow0\right)  =0$, we
find the constraint $\tau_{0}=a_{1}^{-1}a_{2}\left[  \ln\left(  a_{1}^{-1}%
a_{2}\right)  -1\right]$.

\begin{figure}[ptb]
\includegraphics[height=6.25cm]{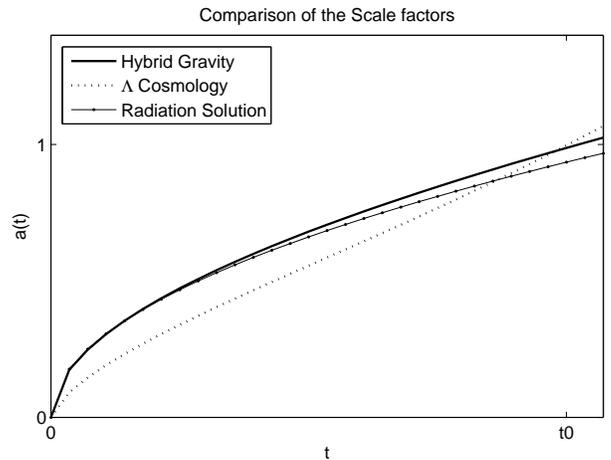}
\caption{Comparison of the scale factor (\ref{Sfhg.01}) with that of 
$\Lambda$CDM-cosmology~$a_{\Lambda}\left(  t\right)  $. and the radiation
solution $a_{r}\left(  t\right)  =a_{0r}\sqrt{t}$ where $t_{0}$ is the present
time, $a_{\Lambda}\left(  t_{0}\right)  =1$. For the solution (\ref{Sfhg.01}) of the Hybrid
Gravity, we set $\left \vert a_{1}\right\vert >1$. }%
\label{fig1}%
\end{figure}

In Figure \ref{fig1} we compare the behavior of the scale factor
(\ref{Sfhg.01}) with that of the standard $\Lambda$CDM-cosmology and the radiation
solution. It can be observed that, in the early universe, the behavior of the scale
factor (\ref{Sfhg.01}) of the Hybrid Gravity is similar to the radiation
solution. 

\subsubsection{\textbf{The  WDW equation}}

From the Hamiltonian (\ref{An.04}),  we can define the WDW equation (recall that
the dimension of the minisuperspace is two and the minisuperspace is flat), which is given by
\begin{equation}
\label{WDW.01}\Psi_{,xx}+\Psi_{,yy}-2V_{0}y^{4}\Psi=0\,,
\end{equation}
where $\Psi$ is the Wave Function of the Universe \cite{halliwell}.
Following the results in \cite{AnIJGMP}, one finds that Eq. (\ref{WDW.01}) admits Lie point symmetries for the vector fields
\begin{eqnarray}
X_{\Psi}&=&c_{1}\partial_{x}+\left(  c_{2}\Psi+b\left(  x,y\right)  \right)
\partial_{\Psi}, 
     \\
 X_{b}&=&b\left(  x,y\right)  \partial_{\Psi} ,
\end{eqnarray}
where $b\left( x,y\right) $ is a function that  satisfies the WDW Eq. (\ref{WDW.01}).
Therefore we can apply the zero order invariants  to reduce Eq.
(\ref{WDW.01}).

From the Lie point symmetry $X_{\Psi}$, the invariant functions are $\left\{
y,Ye^{\mu x}\right\} $, with $\mu\in \mathbb{C}$ \cite{StephaniB}, hence Eq. (\ref{WDW.01}) reduces to the following
second order ODE
\begin{equation}
Y_{,yy}+\left(  \mu^{2}-2\bar{V}_{0}y^{4}\right)  Y=0.
\end{equation}
This equation is the one-dimensional time-dependent oscillator and admits
eight Lie point symmetries \cite{LeachOSc} which are all Type II hidden
symmetries \cite{Abraham,TypeII}. 
Therefore we have that
\[
Y\left(  y\right)  =y_{1}e^{w\left(  y\right)  }+y_{2}e^{-w\left(  y\right)  },
\]
where 
\begin{equation}
w\left(  y\right)  =\frac{\sqrt{2}}{2}\int\sqrt{\left(  2\bar{V}%
_{0}y^{4}-\mu^{2}\right)  }dy.
\end{equation}
Hence, the invariant solution of the WDW equation (\ref{WDW.01}) is finally given by
\begin{equation}
\Psi\left(  x,y\right)  =\sum_{\mu}\left[  y_{1}e^{\mu x+w\left(  y\right)
}+y_{2}e^{\mu x-w\left(  y\right)  }\right]\,.
\end{equation}

\subsection{The case of the potential $V\left(  \phi\right)  =V_{0}\left(  1+\phi\right)
^{2}\exp\left(  \frac{6}{V_{1}}\arctan\sqrt{\phi}\right)  $}

\subsubsection{\textbf{  Lagrangian,  Hamiltonian, and  field equations}}

As before, for the considered potential, we apply the coordinate transformations
\begin{eqnarray}
a&=&\frac{1}{\sqrt{12}}\frac{e^{u}}{\sqrt{\tan^{2}\left(  v-V_{1}u\right)  +1}}, 
    \\ 
\phi &=&\tan^{2}\left(  v-V_{1}u\right),
\end{eqnarray}
then the Lagrangian (\ref{FRW.02})  becomes%
\begin{eqnarray}
{\cal L}\left(  r,\theta,r^{\prime},\theta^{\prime}\right)  &=& \frac{1}{2}e^{2u}\left[
\left(  1+V_{1}^{2}\right)  u^{\prime2}-2V_{1}u^{\prime}v^{\prime}+v^{\prime
2}\right]  
    \nonumber \\
&&+\bar{V}_{0}e^{-2u}e^{\frac{6}{V_{1}}v} , \label{An.09}
\end{eqnarray}
where $\bar{V}_{0}=V_{0}/144$. The Hamiltonian of the system is%
\begin{equation}
\tilde{H}=\frac{1}{2}e^{-2u}\left[  p_{u}^{2}+2V_{1}p_{u}p_{v}+\left(
1+V_{1}^{2}\right)  p_{v}^{2}\right]  -\bar{V}_{0}e^{-2u}e^{\frac{6}{V_{1}}v} .
\label{An.10}%
\end{equation}
The Hamilton equations are 
\begin{eqnarray}
u^{\prime}&=&e^{-2u}\left(  p_{u}+V_{1}p_{v}\right) ,
   \\
v^{\prime}&=&e^{-2u}\left(  V_{1}p_{u}+\left(  1+V_{1}^{2}\right)  p_{v}\right) ,
\end{eqnarray}
\begin{equation}
p_{v}^{\prime}=\frac{6\bar{V}_{0}}{V_{1}}e^{-2u}e^{\frac{6}{V_{1}}v} ,
\end{equation}%
\begin{equation}
p_{u}^{\prime}=e^{-2u}\left(  p_{u}^{2}+2V_{1}p_{u}p_{v}+\left(  1+V_{1}%
^{2}\right)  p_{v}^{2}\right)  -2\bar{V}_{0}e^{-2u}e^{\frac{6}{V_{1}}v} ,
\end{equation}
respectively, and the Hamiltonian constraint provides $\tilde{H}=0$. 
Furthermore, from the Hamilton-Jacobi equation for Eq. (\ref{An.10}), we have the action%
\begin{eqnarray}
S\left(  u,v\right)  &=& \frac{c_{1}}{1+V_{1}^{2}}u-c_{1}\frac{V_{1}}{1+V_{1}%
^{2}}v
     \nonumber \\
&& \hspace{-1cm} +\frac{V_{1}}{3\left(  1+V_{1}^{2}\right)  }\left(  S_{1}\left(
v\right)  -c_{1}\arctan\frac{S_{1}\left(  v\right)  }{c_{1}}\right) ,
\end{eqnarray}
where%
\begin{equation}
S_{1}\left(  v\right)  =2\left(  1+V_{1}^{2}\right)  V_{0}e^{\frac{6}{V_{1}}%
v}-c_{1}^{2}.
\end{equation}
Finally, the reduced dynamical system has the form%
\begin{equation}
e^{2u}u^{\prime}=c_{1}\frac{1-V_{1}^{2}}{1+V_{1}^{2}}+\frac{V_{1}}{1+V_{1}%
^{2}}S_{1}\left(  v\right) , \label{Rds.01}%
\end{equation}%
\begin{equation}
e^{2u}v^{\prime}=S_{1}\left(  v\right)  -c_{1}\frac{V_{1}^{3}}{1+V_{1}^{2}},
\label{Rds.02}%
\end{equation}
which is a system of two nonlinear first order differential equations. However
in order to simplify the reduced system of Eqs. (\ref{Rds.01}) and (\ref{Rds.02}) and
to write the exact solution of the field equations, we apply a second conformal
transformation~$ds=e^{2u}d\tau,$ so that the dynamical system becomes%
\begin{align*}
\frac{du}{ds}  &  =C_{1}+C_{2}+C_{3}e^{\frac{6}{V_{1}}v}  ,\\
\frac{dv}{ds}  &  =C_{4}e^{\frac{6}{V_{1}}v}+C_{5},
\end{align*}
where the constants  are $C_{1..5}=C_{1..5}\left(  V_{0},V_{1}%
,c_{1}^{2}\right) $. The solution of the system can be written as
follows%
\begin{eqnarray}
u\left(  s\right)   &  =& -\frac{V_{1}C_{3}}{6C_{4}}\ln\left\{ C_{4}\left[
\exp\left(  \frac{6C_{5}}{V_{1}}\left(  s+I_{0}\right)  -u_{1}\right)
-1\right]  \right\}  
    \nonumber   \\    
&&    +\left(  C_{1}+C_{2}\right)  s+u_{2}  ,
     \\
v\left(  s\right)   & = &  -\frac{V_{1}}{6}%
\ln\left\{  \frac{1}{C_{5}}\left[ 1-\exp\left(  \frac{6C_{5}}{V_{1}}\left(
s+I_{0}\right)  -u_{1}\right)  \right]  \right\}
    \nonumber  \\
 && +    C_{5}\left(  s+u_{1}\right)  .
\end{eqnarray}
We note that the second conformal transformation $\tau
\rightarrow s$ is of the form $ds=\bar{N}\left(
a,\phi\right)  d\tau$.

\subsubsection{\textbf{The WDW equation}}

Also in this case, from the Hamiltonian (\ref{An.10}), we can define the WDW equation
\begin{equation}
\Psi_{,uu}+2V_{1}\Psi_{,uv}+\left(  1+V_{1}^{2}\right)  \Psi_{,vv}-4\bar
{V}_{0}e^{\frac{6}{V_{1}}v}\Psi=0\, .  \label{WDW.04}%
\end{equation}

Following \cite{AnIJGMP}, we find that Eq. (\ref{WDW.04}) admits, as Lie
point symmetries, the following vector fields%
\[
X_{1}=\partial_{u}, \qquad   X_{\Psi}=\Psi\partial_{\Psi}, \qquad  X_{b}=b\left(
u,v\right)  \partial_{\Psi} ,
\]%
\begin{eqnarray}
X_{2}  &=& e^{-\frac{3v}{V_{1}}}\left[  \cos\left(  V_{C}u\right)  \cos\left(
3v\right)  +\sin\left(  V_{C}u\right)  \sin\left(  3v\right)  \right]
\partial_{u}
   \nonumber   \\
&&  +e^{-\frac{3v}{V_{1}}}\Big[  \left(  V_{1}\cos\left(  3v\right)
-\sin\left(  3v\right)  \right)  \cos\left(  V_{C}u\right)  
  \nonumber  \\
&&+\left(
\cos\left(  3v\right)  +V_{1}\sin\left(  3v\right)  \right)  \sin\left(
V_{C}u\right)  \Big]  \partial_{v} ,
\end{eqnarray}%
\begin{eqnarray}
X_{3}  &=& e^{-\frac{3v}{V_{1}}}\left[  \cos\left(  V_{C}u\right)  \sin\left(
3v\right)  +\sin\left(  V_{C}u\right)  \cos\left(  3v\right)  \right]
\partial_{u}
    \nonumber  \\
&&  +e^{-\frac{3v}{V_{1}}}\Big[  \left(  \cos\left(  3v\right)  +V_{1}%
\sin\left(  3v\right)  \right)  \cos\left(  V_{C}u\right)  
     \nonumber   \\
&&+\left(\sin\left(  3v\right)  -V_{1}\cos\left(  3v\right)  \right)  
\sin\left( V_{C}\right)  \Big]  \partial_{v} ,
\end{eqnarray}
where $V_{C}=3\left(  1+V_{1}^{2}\right) /V_{1}$. We note that only the symmetry vector $X_{1}=\partial_{u}$ is the generator of the Noether symmetry for the  Lagrangian (\ref{An.09}).

We apply the invariant  symmetry vector $X=X_{1}+\mu\Psi\partial_{\Psi}%
$, where the invariants are $\left\{  v,Ye^{\mu u}\right\}  $. Therefore Eq.
(\ref{WDW.04}) becomes%
\begin{equation}
\left(  1+V_{1}^{2}\right)  Y_{,vv}+2\mu V_{1} Y_{,v}+\left(  \mu^{2}-4\bar
{V}_{0}e^{\frac{6}{V_{1}}v}\right)  Y=0 \,.  \label{WDW.05}%
\end{equation}
This equation describes a time-dependent damped oscillator and it is well know
that there exists a transformation $\left(  v,Y\right)  \rightarrow\left(
\bar{v},\bar{Y}\right)  $ where it can be written in the form $\bar
{Y}_{,\bar{v}\bar{v}}=0$, since it admits eight Lie point symmetries.

Therefore the solution of Eq. (\ref{WDW.05}) can be expressed in terms of Bessel
functions%
\[
Y\left(  v\right)  =\exp\left(  -3\bar{N}y\right)  \left[  c_{1}J_{\bar{N}%
}\left(  V_{\mu}e^{\frac{3}{V_{1}}v}\right)  +c_{2}Y_{\bar{N}}\left(  V_{\mu
}e^{\frac{3}{V_{1}}v}\right)  \right]\,,
\]
where
\[
\bar{N}=-\frac{V_{1}\mu}{3\left(  1+V_{1}^{2}\right)  },  \qquad  
 V_{\mu}=\frac{2}%
{3}\frac{V_{1}\sqrt{V_{0}}}{\sqrt{1+V_{1}^{2}}}\, i \,.
\]
Finally the invariant solution of Eq. (\ref{WDW.04}) is
\begin{eqnarray}
\Psi\left(  u,v\right)  &=&\sum_{\mu}\exp\left(  \mu u-3\bar{N}v\right)  \Bigg[
c_{1}J_{\bar{N}}\left(  V_{\mu}e^{\frac{3}{V_{1}}v}\right)  
  \nonumber  \\
&&+c_{2}Y_{\bar{N}%
}\left(  V_{\mu}e^{\frac{3}{V_{1}}v}\Bigg)  \right] .
\end{eqnarray}

However, we note that it is possible to apply the other Lie
symmetries, e.g., $X_{2},X_{3}$ or any linear combination, in order to
determine the invariant solution of the WDW Eq. (\ref{WDW.04}).

\section{ Discussion and Conclusions}
\label{Conclusion}

In this work we considered the application of point symmetries in the Hybrid 
Gravity in order to select the $f\left(  \mathcal{R}\right)  $ function and to
find analytical solutions of the field equations and of the WDW equation for
Quantum Cosmology. We showed that, in order to find  non-linear, integrable $f\left(
\mathcal{R}\right) $ models, we have to apply
conformal transformations in the Lagrangian. Conformal
transformations of the forms $d\tau=N\left(  a\right)  dt$~and $d\tau=N\left(
\phi\right)  dt$ allow to achieve the results. In the second case the Lagrangian of the field equations
reduced to that of a Brans-Dicke-like theory with a general coupling function; then
the results from \cite{TsamC02} for scalar-tensor models can be applied. For
the first conformal transformation we find two cases of the $f\left(
\mathcal{R}\right) $ function where the field equations admit Noether
symmetries. For each case, we transform the field equation by means of  normal
coordinates to simplify the dynamical system and write  exact solutions. Furthermore, we
have  written the WDW equation for the 2-dimensional minisuperspace. 
The Lie point symmetries for the WDW equations can be determined and applied  in order 
to find invariant solutions of the WDW equations. 

However,
it is possible to apply another more general conformal transformation of the
form $d\tau=N\left(  a,\phi\right)  dt$.  If we do not consider
matter, the field equations are always conformally invariant \cite{TsamC01}.
Furthermore, the WDW equation is also conformally invariant, hence the
solutions that we obtained hold for any frame \cite{CapHD,AnIJGMP}.

It is  interesting to stress that, in the case of the power law potential $V\left(
\phi\right)  =V_{0}\left(  \sqrt{\phi}+V_{1}\right)  ^{4}$,  the Hubble function
$H^{2}\left(  z\right)  $ is a fourth-order polynomial with non-vanishing
coefficients. More specifically, every power law term of $\sqrt{\phi}$ in the potential
produces a corresponding fluid in the model. However, we recall that, in the
case where $V_{1}=0$, the solution of the scale factor is the radiation
solution as it has to be for conformally invariant solutions.
Finally, we have to stress  that the only power law Hybrid Gravity which
admits Noether symmetries is  $f\left(  \mathcal{R}\right)  \varpropto
\mathcal{R}^{2}$. This result  is different for   $f\left(  R\right)  $-metric gravity and   $f\left(
\mathcal{R}\right)  $-Palatini gravity where the power law functions which
admits Noether symmetries are $f\left(  R\right)  =R^{n}$~and~$f\left(
\mathcal{R}\right)  $ $=\mathcal{R}^{n}$ \cite{PTB,PalNEB}.
In a forthcoming study, we will face the problem in a more general way in order to achieve more general forms of Hybrid Gravity than simple power-law functions.

As a final remark, it is worth stressing that Hybrid Gravity seems  capable, in principle,    of recovering the various cosmological eras starting from inflation up to dark energy (see also  \cite{Capozziello:2013uya}). However, further refined studies are necessary in order to achieve  in a self-consistent way the full cosmic dynamics.

\acknowledgments{AB and AW acknowledge financial  support from the NCN project 
DEC-2013/09/B/ST2/03455.
SC,  MDL, AP, and MP acknowledge financial  support of INFN (initiative specifiche  QGSKY, QNP, and TEONGRAV ).
FSNL acknowledges financial  support of the Funda\c{c}\~{a}o para a
Ci\^{e}ncia e Tecnologia through an Investigador FCT Research contract, with
reference IF/00859/2012, funded by FCT/MCTES (Portugal), and grants
CERN/FP/123618/2011 and EXPL/FIS-AST/1608/2013. }

\end{document}